\documentclass[aps,prd,eqsecnum,nofootinbib,showpacs,twocolumn]
{revtex4-1}
\usepackage{amsmath,amssymb,bm}
\usepackage[dvips]{graphicx}
\usepackage{color}
\newcommand{\be}{\begin{equation}}
\newcommand{\ee}{\end{equation}}
\newcommand{\bea}{\begin{eqnarray}}
\newcommand{\eea}{\end{eqnarray}}

\newcommand{\eq}[1]{Eq.~(\ref{eq:#1})}
\newcommand{\sect}[1]{Sec.~\ref{sec:#1}}
\newcommand{\appen}[1]{Appendix~\ref{sec:#1}}

\newcommand{\del}{\partial}

\bmdefine{\bmg}{{\bm{g}}}
\bmdefine{\bmk}{{\bm{k}}}
\bmdefine{\bmq}{{\bm{q}}}
\bmdefine{\bmx}{{\bm{x}}}
\bmdefine{\bmA}{{\bm{A}}}
\bmdefine{\bmH}{{\bm{H}}}
\bmdefine{\bmS}{{\bm{S}}}
\bmdefine{\bmphi}{{\bm{\phi}}}
\bmdefine{\bmpsi}{{\bm{\psi}}}
\bmdefine{\bmPhi}{{\bm{\Phi}}}
\bmdefine{\bmPsi}{{\bm{\Psi}}}
\bmdefine{\bmnabla}{{\bm{\nabla}}}
\newcommand{\calA}{\mathcal{A}}

\newcommand{\calF}{\mathcal{F}}

\newcommand{\calL}{\mathcal{L}}
\newcommand{\calM}{\mathcal{M}}
\newcommand{\calO}{\mathcal{O}}

\newcommand{\calR}{\mathcal{R}}

\bmdefine{\bmcalA}{{\bm{\calA}}}
\newcommand{\tilA}{\Tilde{A}}
\newcommand{\tilF}{\Tilde{F}}
\newcommand{\tilG}{\Tilde{G}}
\newcommand{\tilJ}{\Tilde{J}}
\newcommand{\tilK}{\Tilde{K}}
\newcommand{\tilS}{\Tilde{S}}

\newcommand{\tilrho}{\Tilde{\rho}}

\newcommand{\tcA}{\Tilde{\calA}}

\newcommand{\tct}{{\text{ct}}}
\newcommand{\tint}{{\text{int}}}
\newcommand{\tBF}{{\text{BF}}}
\newcommand{\odiff}[2]{ \frac{d #1}{d #2} }

\newcommand{\pdiff}[2]{ \frac{\partial #1}{\partial #2} }

\newcommand{\ket}[1]{\left\vert\,#1\,\right\rangle}
\newcommand{\bra}[1]{\left\langle\,#1\,\right\vert}
\newcommand{\Exp}[1]{{\langle\, #1\, \rangle}}
\newcommand{\varExp}[1]{\left\langle~#1~\right\rangle}
\newcommand{\Tr}{\text{Tr}}
\begin{document}

\title{Two pieces of folklore in the AdS/CFT duality}
\author{Kengo Maeda}
\email{maeda302@sic.shibaura-it.ac.jp}
\affiliation{Department of Engineering,
Shibaura Institute of Technology, Saitama, 330-8570, Japan}
\author{Makoto Natsuume}
\email{makoto.natsuume@kek.jp}
\affiliation{KEK Theory Center, Institute of Particle and Nuclear Studies, 
High Energy Accelerator Research Organization,
Tsukuba, Ibaraki, 305-0801, Japan}
\author{Takashi Okamura}
\email{tokamura@kwansei.ac.jp}
\affiliation{Department of Physics, Kwansei Gakuin University,
Sanda, Hyogo, 669-1337, Japan}
\date{\today}
\begin{abstract}
In the AdS/CFT duality, it is often said that a local symmetry in a bulk theory corresponds to a global symmetry in the corresponding boundary theory, but the global symmetry can become local when one couples with an external source. As a result, the GKP-Witten relation gives a response function instead of a Green function. We explore this point in detail using the example of holographic superconductors. We point out that these points play a crucial role in interpreting the holographic London equation properly.
\end{abstract}
\pacs{11.25.Tq, 74.20.-z} 

\maketitle

\section{Introduction and Summary} \label{sec:intro}


There are often quoted pieces of ``folklore" in the AdS/CFT (anti-de~Sitter/conformal field theory) duality \cite{Maldacena:1997re,Witten:1998qj,Witten:1998zw,Gubser:1998bc}. Two popular ones are \begin{enumerate}

\item[(i)]
``A local symmetry in a bulk theory corresponds to a global symmetry in the corresponding boundary theory," {\it e.g.}, 
$U(1)$ symmetry in holographic superconductors.

\item[(ii)]
``The GKP (Gubser-Klebanov-Polyakov)-Witten relation \cite{Maldacena:1997re,Witten:1998qj,Gubser:1998bc} gives the Green functions (correlation functions)."

\end{enumerate}
However, they are not entirely true when one couples the boundary theory with an external source. 

As an example, consider a bulk Maxwell field. Then, a current in the boundary theory is coupled with the external source which is given by the bulk Maxwell field. Note that the Maxwell field acts only as a source in the boundary theory and there is no dynamical photon in the field theory side. (See footnote~\ref{fnote:dynamical} for a subtlety.) In this case,
\begin{enumerate}

\item[(i')]
In the presence of the coupling with the external source, one can promote the global symmetry to a local symmetry by assigning a local transformation of the external source. We call it the ``background local $U(1)$ symmetry" (\sect{first}).

\item[(ii')]
When the first folklore fails due to the background local $U(1)$ symmetry, the second folklore is not true either in general. The AdS computations are naturally associated with the {\it response function} not with the Green function, and they differ in general (\sect{second}).

\end{enumerate}

The reader may wonder if the statement~(i') is just a matter of convention. After all, the background local symmetry is not a local symmetry in the usual sense. Whether one calls it a ``local symmetry" is not really an issue. The point is that the statement~(i') implies that the boundary current contains the external source. 
This leads to an important physical consequence, which is the statement~(ii').%
\footnote{The statement~(i') may not be a necessary condition for the statement~(ii') however. See the end of \sect{first}. }
The response function contains additional terms which come from the explicit dependence of the current on the source. This is the origin of the difference between the response function and the Green function. 


The ``breakdown" of these folklore statements is not limited to the AdS/CFT duality; it equally applies to a field theory. In fact, we will use a simple field theory example to illustrate the points. Thus, these points are probably well-known to some experts. 

However, the situation is more subtle in the AdS/CFT duality. First, the boundary field theory description is often unavailable in holographic applications such as holographic superconductors \cite{Gubser:2008px,Hartnoll:2008kx,Hartnoll:2008vx}. (See, {\it e.g.}, Refs.~\cite{Hartnoll:2009sz,Herzog:2009xv,Horowitz:2010gk} for reviews.) So, it may be worthwhile to emphasize these points. Second, since the boundary description is not available, one had better show these points without relying on the boundary description. We argue from the bulk theory point of view that the statements (i') and (ii') must hold. There is a drawback of such an approach though: our argument is somewhat indirect.

In particular, these points play a crucial role to properly interpret the ``holographic London equation," which is an example of response functions. The difference between the response function and the Green function is particularly sharp in this case since they differ by a sign.%
\footnote{
The traditional {\it sign convention} of the response function differs from the Green function (see \appen{conventions}). Our argument in this paper focuses on the sign of these functions, so the traditional convention is somewhat confusing. We define the response function in such a way that it has the same sign as the Green function.
}


More precisely, according to the holographic London equation, the response function is positive, but in general (the diagonal part of) the Green function must be negative from the spectral representation (K\"{a}llen-Lehmann representation). Thus, one can explain the sign of the holographic London equation only if one takes their difference into account. 

We also would like to point out the following issue. 
The difference between these functions should resolve the above sign problem,
and this is indeed the case for standard superconductors (\sect{second}). 
But it has never been shown for holographic superconductors that the Green function itself is negative-definite. It is desirable to compute the Green function itself in the holographic superconductors and desirable to show this point explicitly for internal consistency. 

In this paper, we focus on the bulk local $U(1)$ symmetry as an explicit example, but a similar statement holds for the diffeomorphism invariance and the local supersymmetry in a bulk theory. 

The plan of this paper is as follows. 
We describe elementary facts about the London equation (and its relation to a response function) and its holographic counterpart in \sect{London}. We show the breakdown of the first folklore and the second folklore in Secs.~\ref{sec:first} and \ref{sec:second}, respectively. Our main focus is the holographic London equation, so its validity is particularly important. We show the holographic London equation rather generically in \appen{holo^London_eq}.


In the rest of this introduction, let us briefly describe holographic superconductors. (See \appen{preliminaries} for more details.) Holographic superconductors are described by Einstein-Maxwell-complex scalar system in asymptotically $(p+2)$-dimensional AdS spacetime:
\begin{align}
   \frac{\calL_{\text{full}}}{ \sqrt{- g} }
  &= R - 2 \Lambda - \frac{ F^{MN} F_{MN} }{4}
  - \left\vert D \Psi \right\vert^2 - V\left( | \Psi |^2 \right)~,
\label{eq:full_action}
\end{align}
where 
%
\begin{align}
  & F_{MN} = 2\, \partial_{[M} A_{N]}~,
& & D_M := \nabla_M - i e A_M~,
\label{eq:def-covariant_deri} \\
  & \Lambda
  = - \frac{p (p + 1)}{2\, l^2}~,
& & V = m^2 | \Psi |^2~.
\label{eq:def-potential_V}
\end{align}
Here, capital Latin indices $M, N, \ldots$ run through bulk spacetime coordinates $(t, \bmx, u)$, where $(t,\bmx)=(t,x_i)$ are the boundary coordinates and $u$ is the AdS radial coordinate. Greek indices $\mu, \nu, \ldots$ run though only the boundary coordinates.

We consider the matter fields which behave asymptotically $u\rightarrow 0$ as 
\begin{subequations}
\begin{align}
  & \bmA_t(u)
  \sim \mu - \frac{l\, \rho}{(p - 1) (l\, \zeta)^{p-1}}\, u^{p-1}~,
\label{eq:bg-asymptotic_behavior_i} \\
  & \bmPsi(u)
  \sim \psi^{(-)} u^{\Delta_-} + \psi^{(+)} u^{\Delta_+} \qquad (\Delta_-<\Delta_+),
\end{align}
\end{subequations}
where boldface letters are used for background values. Also, $\rho$ is the charge density in the boundary theory, $\mu$ is the chemical potential associated with the charge density, and $\zeta$ is related to the Hawking temperature $T$ as
$4 \pi\, T = (p + 1)\, \zeta$. According to the standard AdS/CFT dictionary, $\psi^{(+)}$ represents the expectation value of a field theory operator $\calO$, so it represents a condensate. 
As is clear from \eq{bg-asymptotic_behavior_i}, the bulk fields act only as external sources of boundary operators in the AdS/CFT duality. Thus, there is no dynamical photon in the boundary theory.%
\footnote{Actually, this is another folklore which may not be true in general. In low spatial dimensions ($p \leq 2$), the bulk Maxwell field can induce a theory with a dynamical gauge field (see, {\it e.g.}, Refs.~\cite{Witten:2003ya,Marolf:2006nd,Imamura:2008ji}). See Ref.~\cite{Domenech:2010nf} for an implementation to holographic superconductors. To exclude this possibility, consider, {\it e.g.}, $p>2$.
\label{fnote:dynamical}
}

The model exhibits a second-order phase transition. Its critical phenomena near the phase transition have been discussed in Ref.~\cite{Maeda:2009wv}. At high temperatures, the scalar field $\Psi$ vanishes and one obtains the standard Reissner-Nordstr\"{o}m-AdS black hole. But at low temperatures, the Reissner-Nordstr\"{o}m-AdS black hole becomes unstable and is replaced by a charged black hole with a scalar ``hair." Furthermore, the low temperature solution has the expected behavior for a superconducting phase, {\it i.e.}, (1) the divergence of the DC conductivity, and (2) an energy gap proportional to the size of the condensate.



\section{Electromagnetic response} \label{sec:London}


\subsection{The London equation as a response function}
\label{sec:EV-response_func}

A superconductor has singular behaviors in the transport properties of the current. But its essence is not in the diverging conductivity but in the Meissner effect which expels a magnetic field. A diverging conductivity also appears in a perfect conductor, but the Meissner effect is unique to superconductors.%
\footnote{
The {\it exclusion} of a magnetic field from entering a superconductor can be explained by perfect conductivity. On the other hand, a magnetic field in an originally normal sample is also {\it expelled} as it is cooled below $T_c$. This cannot be explained by perfect conductivity since it tends to trap flux in. 
}

Phenomenologically, the Meissner effect is a consequence of the London equation and the Maxwell equation. The London equation is given by
\begin{align}
  &\Exp{J_i(\bmx)}
  \sim - \frac{n_s\, e_*^2}{m_*}\, \calA_i(\bmx)~,
\label{eq:London_eq}
\end{align}
where $e_*$ and $m_*$ represent the effective charge and the effective mass of Cooper pairs, and $n_s$ represents the superfluid density. Note that the London equation is not gauge-invariant. The London equation is valid in  the London gauge 
\be
\partial_i \calA^i = 0~. 
\ee
Combined the London equation with the Maxwell equation, a magnetic field decays exponentially inside a superconductor. Note that a dynamical photon is mandatory to have the Meissner effect.

The London equation is an example of a linear response relation: the response of the system ($J_i$) is linearly related to the perturbation ($\calA_i$). They are related by a {\it response function}. 

It is convenient to introduce the response function $K^{ij}$ in order to express the generic electromagnetic response of a superconductor: 
\begin{subequations}
\label{eq:Kubo_formula}
\begin{align}
  & \delta \Exp{J^{i}(x)}
  = - \int^\infty_{- \infty} d^{p+1}x'~
  K^{ij}(x - x'\,)\, \delta \calA_{j}(x'\,)~,
\label{eq:config-Kubo_formula} \\
  &\Leftrightarrow \hspace{0.2truecm}
   \delta \Exp{\tilJ^{i}(\omega, \bmk)}
  = - \tilK^{ij}(\omega, \bmk)\, \delta \tcA_{j}(\omega, \bmk)~,
\label{eq:FT-Kubo_formula}
\end{align}
\end{subequations}
where ``~$\delta$~" denotes the deviation from the background value, 
$x = (t, \bmx)$, and  \lq\lq\, $\Tilde{~}$~\rq\rq\, means Fourier-transformed quantities, {\it e.g.},
\begin{align}
  & \tilF(\omega, \bmk)
  := \int^\infty_{- \infty} dt\, d\bmx~
  e^{i \omega t - i \bmk \cdot \bmx}~F(t, \bmx)~.
\label{eq:def-FT}
\end{align}
Since our interest is in the response to a static source, we will consider the $\omega\rightarrow0$ limit of $\tilK^{ij}$ which will be called as the ``static response function." Note that the generic linear response relation (\ref{eq:Kubo_formula}) is a nonlocal expression whereas the London equation is a local expression. This is because the London equation is a phenomenological equation so the long-wavelength limit is implicitly assumed. The nonlocal extension of the London equation is known as the Pippard equation. (It is in the same spirit as the second-order hydrodynamics. See, {\it e.g.}, Ref.~\cite{Natsuume:2008ha} for a review.)


Let us rewrite the London equation in terms of $\tilK^{ij}$. It is convenient to use the tensor decomposition:
\begin{align*}
  & \tilK^{ij}(\bmk)
  =: \frac{k^i\, k^j}{| \bmk^2 |}\, \tilK_{L}(\bmk)
  + \left( \delta^{ij} - \frac{k^i\, k^j}{| \bmk^2 |} \right)\,
    \tilK_{T}(\bmk)~.
\end{align*}
If one takes the frame $k^i=(0,\cdots, 0, k)$, 
$\tilK^{ij}={\rm diag}(\tilK_{T},\cdots, \tilK_{T},\tilK_{L})$. 
The London equation (\ref{eq:London_eq}) employs the London gauge $\partial_i \calA^i = 0$.
Thus, the London equation holds if the transverse part of $\tilK^{ij}$ is positive-definite in the long-wavelength limit:
\begin{align}
  & \lim_{\bmk \to 0} \tilK_{T}(\bmk)
  = C > 0~,
\label{eq:guess-holographic_London_eq}
\end{align}
where $C:=n_s e_*^2/m_*>0$.

\subsection{The holographic London equation}

For holographic superconductors, there is no dynamical photon in the field theory.
Thus, the Meissner effect does not arise, and a magnetic field can penetrate superconductors. Therefore, holographic superconductors are extreme type II superconductors just like a superfluid \cite{Hartnoll:2008kx,Herzog:2008he,Basu:2008st}. In type II superconductors, the penetration of the magnetic field arises by forming vortices. The vortex solutions have been constructed for holographic superconductors \cite{Albash:2009ix,Albash:2009iq,Montull:2009fe,Maeda:2009vf}. 

Even though the Meissner effect does not arise, the holographic London equation must hold. The London equation is just the response of the current under the external source. Whether photon is dynamical or not should be irrelevant to the response itself. In order to show that a holographic superconductor is really a superconductor, it is important to check the holographic London equation.

The current expectation value of the boundary theory is evaluated from the GKP-Witten relation:
\begin{subequations}
\bea
%
e^{W[\calA_\mu]}
&=& e^{-S_\text{os}[\calA_\mu]}~,
\label{eq:GKP-W} \\
\calA_\mu &:=& A_\mu \big\vert_{u=0}~.
\eea
\end{subequations}
The left-hand side is the generating function of the boundary theory located at $u=0$, and the right-hand side is the generating function of the bulk theory with the on-shell bulk action $S_\text{os}$. The boundary value of the bulk $U(1)$ field $A_\mu$ is denoted as $\calA_\mu$.
This is the standard Euclidean prescription of the AdS/CFT duality, not the Lorentzian prescription in Ref.~\cite{Son:2002sd}. The Lorentzian prescription is often used to study dynamics.
Our interest in this paper is the static response function, so it is enough to use the GKP-Witten relation.

From the GKP-Witten relation, the current expectation value is given by
\begin{subequations}
\label{eq:R_current-exp_value}
\bea
\Exp{\tilJ^{i}(\bmk)}
  &:=& (2 \pi)^{p+1} \frac{\delta W[\tcA]}{\delta \tcA_{i}(-\bmk)} 
\\
  &=& - (2 \pi)^{p+1}\,
     \frac{ \delta S_{\text{os}} }
          { \delta \tcA_{i}(-\bmk) }\, \bigg\vert_{u=0}~.
\eea
\end{subequations}
%
%
Then, the static response function is given by
%
\begin{subequations}
\label{eq:R_current-linear_response}
\begin{align}
  & \delta \Exp{\tilJ^{i}(\bmk)}
  = - \tilK^{ij}(\bmk)\, \delta \tcA_{j}(\bmk)~,
\label{eq:R_current-Kubo_formula} \\
  & \tilK^{ij}(\bmk)
  := \left. (2 \pi)^{p+1}\,
  \frac{ \delta^2 S_{\text{os}} }
      { \delta \tcA_i(- \bmk)\, \delta \tcA_j(\bmk) }
  \, \right\vert_{u=0}~.
\label{eq:def-R_current-response_func}
\end{align}
\end{subequations}
%
The issue is whether \eq{def-R_current-response_func} behaves like the London equation (\ref{eq:guess-holographic_London_eq}) in the long-wavelength limit.

The holographic London equation has been shown for the Maxwell-complex scalar system on the four-dimensional Schwarzschild-AdS background 
($\text{SAdS}_4$) in the ``probe limit" \cite{Maeda:2008ir,Hartnoll:2008kx}. In \appen{holo^London_eq}, we show the holographic London equation rather generically in the probe limit.

The reader may recall Weinberg's derivation of the London equation \cite{Weinberg:1986cq}. He derived the London equation as a consequence of local $U(1)$ spontaneous symmetry breaking from the point of view of effective theory. The argument holds even when the gauge field is nondynamical, {\it i.e.}, the argument uses only the background local $U(1)$ symmetry. Then, Weinberg's argument with our argument in \sect{first} (the existence of the background local $U(1)$ symmetry in the boundary theory) immediately implies the validity of the holographic London equation. 

We will not take the path though. In a sense, our standpoint is somewhat opposite to his argument. We use the holographic London equation itself to argue the existence of the background local $U(1)$ symmetry. Also, one emphasis in this paper is the difference between the Green function and the response function in the AdS/CFT duality. We will use the spectral representation of the Green function, so we argue the holographic London equation from a microscopic point of view.

\section{Two pieces of folklore}
\label{sec:folklores}

\subsection{First folklore}
\label{sec:first}

In the AdS/CFT duality, one often says that a local gauge symmetry in a bulk theory corresponds to a global symmetry in the corresponding boundary theory. This is the first folklore we discuss. When one couples with an external source, the global symmetry can become local if one allows a local transformation of the external source. 

It is often convenient to work in the gauge $A_u = 0$ for the bulk $U(1)$ gauge field. This gauge condition does not completely fix the gauge, and there is a residual gauge transformation which leaves the gauge condition $A_u = 0$ invariant: 
\begin{subequations}
\label{eq:residual_gauge}
\begin{align}
   A_M(t, \bmx, u)
  &\to A_M(t, \bmx, u) + \partial_M \Lambda(t, \bmx)~,
\label{eq:residual_gauge-A} \\
   \Psi(t, \bmx, u)
  &\to e^{i e \Lambda(t, \bmx)}\, \Psi(t, \bmx, u)~.
\label{eq:residual_gauge-Psi}
\end{align}
\end{subequations}
Then, the gauge transformation (\ref{eq:residual_gauge}) acts on the source $\calA_\mu$ of the $U(1)$ current and on the operator expectation value $\Exp{\calO}$ dual to $\Psi$ as
\begin{subequations}
\label{eq:local_U1}
\begin{align}
   \calA_\mu(t, \bmx)
  &\to \calA_\mu(t, \bmx) + \partial_\mu \Lambda(t, \bmx)~,
\label{eq:local_U1-A} \\
   \Exp{\calO(t, \bmx)}
  &\to e^{i e \Lambda(t, \bmx)}\, \Exp{\calO(t, \bmx)}~.
\label{eq:local_U1-calO_exp}
\end{align}
\end{subequations}
This is a local $U(1)$ transformation of the dual field theory, in the sense that one transforms the external source $\calA_\mu$. Such a transformation of an external source is often discussed in a field theory, {\it e.g.}, the background field method.%
\footnote{
Even though the $U(1)$ symmetry is gauged, $\calA_\mu$ is an external source and not an operator of the dual field theory.
}
%


Now, the 
current expectation value is invariant under the background local $U(1)$ transformation (\ref{eq:local_U1}). From \eq{R_current-exp_value}, one gets
%
\begin{align}
   \Exp{J^\mu}
  &= - \frac{ \delta S_{\text{os}} }{ \delta \calA_{\mu} }
  \, \bigg\vert_{u=0}
  = \sqrt{g}\, F^{\mu\nu}\, (du)_\nu \, \Big\vert_{u=0}~.
\label{eq:R_current-GKP_Witten-config}
\end{align}
Here, we assume that there is no contribution from counterterms (see \appen{counter_term}). Because the right-hand side of \eq{R_current-GKP_Witten-config} is invariant under the background local $U(1)$ transformation (\ref{eq:residual_gauge}), $\Exp{J^\mu}$ is also invariant under the transformation. 

Suppose that the transformation (\ref{eq:local_U1-calO_exp}) holds not only for the expectation values but also for the operators. Also, it is natural to assume that the current has the contribution from the scalar condensate $\calO$. Then, one must make the vector quantity $\Exp{J^\mu}$ from the scalar operator $\calO$. Because $\Exp{J^\mu}$ is invariant under the background local $U(1)$ symmetry, one must use the gauge covariant derivative 
\begin{align}
  & \left( \partial_\mu - i\, e\, \calA_\mu \right) \calO~.
\label{eq:gauge_coupling}
\end{align}
The point is that the $U(1)$ current $J^\mu$ contains the external source $\calA_\mu$. This explains the behavior of the static response function we will see in \sect{second}.

Note that the bulk symmetry constrains the boundary theory. As a simple example, suppose that the boundary theory is given by a complex scalar field $\phi$: 
\be
{\cal L} = |\partial_\mu \phi|^2 + V( |\phi | )~.
\ee
When one couples with the external source $ \calA_\mu$, the bulk $U(1)$ symmetry implies that the boundary theory takes the form, {\it e.g.},
\be
{\cal L}_\calA = \big\vert\, (\partial_\mu - i e\, \calA_\mu) \phi\, \big\vert^2 + V( |\phi | )~,
\label{eq:scalar_with_gauge}
\ee
not the form
\be
{\cal L}'_\calA = {\cal L} - \hat{J}^\mu \calA_\mu,
\label{eq:scalar_with_global}
\ee
where $\hat{J}^\mu=-ie \phi^\dagger \tensor\partial^\mu \phi$. 
While the latter theory still has a conserved current  because of the global $U(1)$ symmetry $\phi \rightarrow e^{ie\Lambda} \phi$, the theory is not gauge-invariant. The difference is the $O(e^2)$ term. A related fact is that one often considers the field theory perturbation of the form $\delta {\cal L} = -J^\mu \calA_\mu$ in the GKP-Witten relation, 
but the perturbation (\ref{eq:scalar_with_gauge}) cannot be written in this form.%
\footnote{
Reference~\cite{Horowitz:2010gk} mentioned that one can view the boundary $U(1)$ symmetry as a ``weakly gauged" symmetry where one takes only $O(e)$ terms into account. There is nothing wrong with the point of view, but the validity of the holographic London equation implies that it makes sense to discuss $O(e^2)$ terms in the action.
}

Let us make further comments to clarify the issue:
\begin{itemize}

\item The actual local symmetry which survives depends on the background one considers. For example, in the presence of a boundary chemical potential $\mu=\calA_0$, 
one has only part of the local symmetry (\ref{eq:local_U1}):
\begin{subequations}
\begin{align}
   \calA_i(t, \bmx)
  &\to \calA_i(t, \bmx) + \partial_i \Lambda(\bmx)~,
\\
   \Exp{\calO(t, \bmx)}
  &\to e^{i e \Lambda(\bmx)}\, \Exp{\calO(t, \bmx)}~.
\end{align}
\label{eq:local_U1_spatial}%
\end{subequations}
We assume the backgrounds which preserve the symmetry (\ref{eq:local_U1_spatial}). Our discussion in this paper still applies as long as one has the background local symmetry of the form (\ref{eq:local_U1_spatial}). This is because our interest is in the holographic London equation, so we focus on the response on the spatial component of the current. The symmetry (\ref{eq:local_U1_spatial}) remains both in the high temperature phase and in the low temperature phase.

\item
The local symmetry in the boundary theory often has an anomaly, {\it e.g.}, ${\cal N}=4$ super-Yang-Mills (SYM) theory. The anomaly in the boundary theory corresponds to the presence of a Chern-Simons term in the bulk theory \cite{Witten:1998qj}. For the $p=3$ case, the Chern-Simons term takes the form
\be
S \sim \int d^5x\, d^{abc} \epsilon^{MNOPQ}A_M^a \del_N A_O^b \del_P A_Q^c~,
\ee
which produces the anomaly:
\be
D^\mu J_\mu^a \sim i d^{abc} \epsilon^{\mu\nu\rho\sigma} \calF_{\mu\nu}^b \calF_{\rho\sigma}^c~,
\ee
where $D^\mu$ is the gauge covariant derivative, small Latin indices $a,b, \ldots$ are R-symmetry $SU(4)_R$ indices, and $d^{abc} = \text{tr}[t^a\{t^b,t^c\}]$. Thus, the local symmetry is broken in general, but if one considers a specific external source, the local symmetry remains. Another way to avoid the anomaly is to choose the gauged $U(1)$ such that $d^{abc}=0$ \cite{CaronHuot:2006te}.

\item
The bulk symmetry itself does not uniquely determine the external electromagnetic coupling of the boundary theory. The coupling (\ref{eq:gauge_coupling}) by the gauge covariant derivative is the simplest, but one would need the other covariant terms in general. This is a typical problem in the AdS/CFT duality. 


\item
As we saw in \eq{gauge_coupling}, the source itself appears in the field theory operator. In other words, the ``operator/field" dictionary is modified once the source is present. We discuss this phenomenon in the context of conserved currents, but this phenomenon itself is much more general than the currents, and various examples are found in the literature.%
\footnote{We thank the anonymous referee for pointing this out to us.}
One example is the chiral condensate operator in the D3-D7 system (see, {\it e.g.}, Ref~\cite{Kobayashi:2006sb}). The theory has fundamental hypermultiplets (``quarks"), which consist of Weyl fermions and scalars. The quark mass acts as the source of the chiral condensate, but the condensate contains a term proportional to the scalar mass. [See Eq.~(A1) of the above reference.] In this case, the appearance of the source is guaranteed by supersymmetry. The superpotential mass term linear in $m$ gives both to the fermion mass term (linear in $m$) and the scalar mass term (quadratic in $m$). As a result, the scalar mass appears in the condensate. In this paper, we focus on conserved currents since we are mainly interested in the consequence of the background $U(1)$ symmetry. 

\end{itemize}

\subsection{Second folklore}
\label{sec:second}

In the AdS/CFT duality, the second derivative of the on-shell action with respect to the bulk $U(1)$ field is often interpreted as the (connected) Green function of the $U(1)$ current:
%
\bea
&& \left. 
  \frac{ \delta^2 S_{\text{os}} }
       { \delta \calA_\mu(x)\, \delta \calA_\nu(x'\,) }
  \, \right\vert_{u=0} 
\nonumber \\
&\rightarrow&   G^{\mu\nu}(x-x'\,)
  = - \Exp{\text{T}_E\, J^\mu(x)\, J^\nu(x'\,)}_c~.
\label{eq:def-R_current-correlator}
\eea
Here, $\text{T}_E$ denotes the Euclidean time-ordering,
\footnote{More precisely, it is the T$^*$-product because we use the path integral formalism.}
and the subscript ``$c$" denotes the connected Green function. This is the second folklore we discuss. This folklore is not true in the presence of the background local $U(1)$ symmetry.

If \eq{def-R_current-correlator} were true, the connected Green function $G^{ij}(\bmx)$ would reduce to the static response function  $K^{ij}(\bmx)$ in the stationary limit from \eq{R_current-linear_response}:
\begin{align}
  & \tilK^{ij}(\bmk) \stackrel{?}{=} \tilG^{ij}(\omega = 0, \bmk)~.
\label{eq:tilK-tilG-I}
\end{align}
But this cannot be true because the holographic London equation implies $\tilK_T(\bmk) > 0$ whereas $\tilG^{ii}(\omega=0, \bmk)<0$ (no sum on $i$).

The negative-definiteness of $\tilG^{ii}(0, \bmk)$ can be seen from the spectral representation (K\"{a}llen-Lehmann representation) of the connected Green function%
\footnote{
Our argument here is a rather formal character partly because the underlying field theory description is not yet available. The integral (\ref{eq:def-G_munu-Lehmann}) often does not converge since the spectral function $\tilde{\rho}^{\mu\nu}$ grows at large $\omega$. On dimensional ground, one expects $\tilde{\rho}^{\mu\nu} \propto \omega^{p-1}$. In a field theory, one regularizes the divergence which can change the sign of Eq.~(\ref{eq:negativity}). In a condensed-matter theory, the divergence does not really matter since the theory typically has a ultraviolet cutoff. We thank Chris Herzog for comments.
}
:
\begin{align}
  & \tilG^{\mu\nu}(\omega_n, \bmk)
  = \int^\infty_{-\infty} \frac{d\omega}{2 \pi}~
  \frac{ \tilrho^{\mu\nu}(\omega, \bmk) }{i\, \omega_n - \omega}~,
\label{eq:def-G_munu-Lehmann}
\end{align}
where $\omega_n = 2 \pi n/\beta$ are the Matsubara frequencies, and 
$\tilrho^{\mu\nu}$ is the spectral function which is the Fourier transformation of 
\footnote{Note that $t$ is the real time.
}
\begin{align}
  & \rho^{\mu\nu}(t, \bmx)
  := \Exp{[ J^\mu(t, \bmx), J^\nu(0, \bm{0}) ]}~.
\label{eq:def-rho_munu}
\end{align}
As seen in \appen{negativity}, 
the spectral function satisfies $\tilrho^{\mu\mu}(\omega, \bmk)/\omega > 0$, which leads to the negative-definitess of the Green function:
\begin{align}
  & \tilG^{ii}(0, \bmk)
  = - \int^\infty_{-\infty} \frac{d\omega}{2 \pi}~
  \frac{\tilrho^{ii}(\omega, \bmk)}{\omega} < 0~.
\label{eq:negativity}
\end{align}
%
%
%

Because the response function and the Green function differ by a sign, \eq{tilK-tilG-I} cannot be true. This contradiction is resolved by noting the fact we saw in \sect{first}. Since the bulk theory has the residual gauge symmetry, the boundary theory has the background local $U(1)$ symmetry. Thus, the $U(1)$ current $J^\mu$ contains the external source $\calA_\mu$. In such a case, the response function can differ from the Green function, and the GKP-Witten relation gives the response function instead of the Green function. Using a simple example, let us illustrate this point.


As an example, 
again consider a complex scalar field $\phi$ which couples to the electromagnetic field $\calA_\mu$: 
\begin{align*}
   S_\calA[\, \phi\, ]
  &= \int d^{p+1}x~
  \bigg( \big\vert\, \left( \partial_\mu - i e\, \calA_\mu \right) \phi\,
         \big\vert^2 + V( |\phi | ) \bigg)~,
\end{align*}
%
The current $J^\mu$ is given by
\begin{align}
  & J^\mu = - \frac{\delta S_\calA}{\delta \calA_\mu}
  = -i\,e\, \phi^\dagger\, \tensor\partial^\mu \phi
  - 2e^2 \calA^\mu\, | \phi\, |^2~.
\label{eq:J_i}
\end{align}
Note that the current contains the electromagnetic field $\calA_\mu$ by the background local $U(1)$ symmetry.

The generating functional
\begin{align}
  & Z[\, \calA\, ] 
  = e^{W[\calA]}
  = \int D\phi^\dagger\, D\phi~e^{- S_{\calA}[\, \phi\, ] }~,
\label{eq:def-Z_calA}
\end{align}
gives the current expectation value as
\begin{align}
  & \Exp{J^\mu}
  = - \varExp{ \frac{\delta S_\calA}{\delta \calA_\mu} }
  = \frac{\delta W[\, \calA\, ]}{\delta \calA_\mu}~.
\label{eq:Exp-J_mu}
\end{align}
%
%
%
Then, the response function $K^{\mu\nu}(x)$ is given by
%
\begin{subequations}
\label{eq:def-response_calK}
\begin{align}
   K^{\mu\nu}(x)
  &:= 
     - \frac{ \delta \Exp{J^\mu(x)} }{ \delta \calA_\nu(0) }
  = 
  - \frac{\delta^2 W[\, \calA\, ]}
         { \delta \calA_\nu(0)\, \delta \calA_\mu(x) }
\nonumber \\
  &= G^{\mu\nu}(x)
  - \varExp{ \frac{\delta J^\mu(x)}{\delta \calA_\nu(0) } }
\label{eq:def-response_calK_munuI} \\
  &= G^{\mu\nu}(x)
  + 2\, e^2\, \delta^{\mu\nu}\, \delta(x)\,
    \Exp{| \phi(x) |^2}~.
\label{eq:def-response_calK_munuII}
\end{align}
\end{subequations}
Here, $G^{\mu\nu}$ is the connected Green function for the current $J^\mu$:
\begin{align}
  & G^{\mu\nu}(x)
  = - \Exp{ \text{T}_E\, J^\mu(x)\, J^\nu(0) }_c~.
\label{eq:def-calG_munu}
\end{align}

Thus, the response function differs from the connected Green function by 
the second term of Eq.~(\ref{eq:def-response_calK_munuII}). Then, the negative definiteness  (\ref{eq:negativity}) of the connected Green function is not reflected in the response function. If the absolute value of the second term of Eq.~(\ref{eq:def-response_calK_munuII}) is bigger than the first term, $\tilK_T > 0$ holds. This is indeed the case for standard superconductors \cite{forster}.

To see this, first note that the second term of \eq{def-response_calK_munuII} is proportional to the total number density $n_{\rm tot}:=\Exp{| \phi(x) |^2}$ not the order parameter squared $\Exp{| \phi(x) |}^2$. Thus, 
this term is nonvanishing both in the high temperature phase and in the low temperature phase. 

The first term of \eq{def-response_calK_munuII} gives a contribution which is proportional to the normal component density $n_n$ \cite{forster}. Then, in the low temperature phase, the difference of these terms gives the superfluid density $n_s = n_{\rm tot} - n_n$, which explains the $n_s$-dependence of the London equation (\ref{eq:London_eq}). On the other hand, in the high temperature phase, there is no superfluid component, and $n_{\rm tot} = n_n$. Thus, these terms make no net contribution. 


Finally, from \eq{def-response_calK_munuII}, $K^{\mu\nu}$ and $G^{\mu\nu}$ differ even if there is no external background, {\it i.e.}, $\bmcalA_\mu=0$.


\section{Discussion} \label{sec:discussion}

In the literature, the terms ``global symmetry" and ``no dynamical photon" are often used interchangeably, but they are not the same. There are three possibilities for a $U(1)$ symmetry:
\begin{enumerate}
\item Local $U(1)$ symmetry with a dynamical photon
\item Background local $U(1)$ symmetry without a dynamical photon
\item Global $U(1)$ symmetry only
\end{enumerate}
For possibility~1, see footnote~\ref{fnote:dynamical}.
As an example of possibility~3, see, {\it e.g.}, \eq{scalar_with_global}. Because of the anomaly, one may classify the ${\cal N}=4$ SYM with a generic $U(1)$ gauge field into this case as well.

Most literature on holographic superconductors does not seem to distinguish possibilities~2 and 3 clearly. We argue that the bulk local $U(1)$ symmetry implies possibility~2 and not possibility~3.%
\footnote{
This does not prohibit the interpretation of holographic superconductors as superfluids \cite{Herzog:2008he}. A bulk local $U(1)$ symmetry itself does not determine the boundary symmetry uniquely. The boundary symmetry depends on the choice of whether or not one allows the large gauge transformation (\ref{eq:residual_gauge}) which does not vanish on the boundary. If one does not allow it, one has only the global $U(1)$ symmetry on the boundary. Namely, to some extent, these three possibilities are a matter of choices one makes when defining the theory. For holographic superconductors, it is natural to allow the large gauge transformation because one would like to discuss their electromagnetic response. If one does not allow the large gauge transformation, the statement~(i) below obviously does not apply. The statement~(ii) is more subtle, but one can probably interpret the holographic London equation along the line of Ref.~\cite{leggett}.
}
We gave two circumstantial evidences for this point of view:
\begin{enumerate}
\item[(i)] 
A bulk local $U(1)$ symmetry contains a background local $U(1)$ symmetry acting on the boundary.
The boundary $U(1)$ current $J^\mu$ is invariant under the background local $U(1)$ transformation (\ref{eq:R_current-GKP_Witten-config}), which suggests that $J^\mu$ contains the external source $\calA_\mu$.

\item[(ii)] 
When $J^\mu$ contains $\calA_\mu$, the second derivative of the GKP-Witten relation does not give the Green function in general. In fact, the second derivative leads to the holographic London equation, which does not satisfy a property for a Green function. Put differently, one cannot explain the sign of the holographic London equation unless one takes the $\calA_\mu$-dependence of the current into account.
\end{enumerate}

The bottom line of the background local $U(1)$ symmetry is that the response function differs from the connected Green function in general and that {\it the AdS computation is naturally associated with the response function and not with the Green function.} This should resolve the above sign problem. 
But it is not clear if the Green function itself really satisfies the desired property in holographic superconductors. It would be interesting to find a way to compute the Green function itself in the AdS/CFT duality and to show this point explicitly.

Finally, this phenomenon, the modification of the operator/field dictionary in the presence of the source, itself is a more generic phenomenon than the conserved currents (see \sect{first}). As a result, there may be examples of (ii) which do not come from the background local symmetry (i). 


\begin{acknowledgments}

We are in debt to various people for their thoughts and opinions on the issue of global symmetry in the AdS/CFT duality. In particular, we would like to thank Yosuke Imamura, Shin Nakamura, Masahiro Ohta, Yasuhiro Sekino, Dam Son, and Shigeki Sugimoto. We would also like to thank Shigeki Sugimoto and Gary Horowitz for their comments on the manuscript.
MN would like to thank IPMU Focus Week ``Condensed matter physics meets high energy physics" (Feb.\ 8-12 2010, IPMU) and Komaba 2010 ``Recent developments in strings and fields" (Feb.\ 13-14 2010, Komaba) where he had the opportunity to discuss this issue with some of above people. 
TO would like thank the Yukawa Institute for Theoretical Physics at Kyoto University. Discussions during the YITP workshop YITP-W-09-04 on ``Development of Quantum Field Theory and String Theory'' were useful to complete this work. 
This research was supported in part by the Grant-in-Aid
for Scientific Research (20540285) from the Ministry of Education,
Culture, Sports, Science and Technology, Japan.
\end{acknowledgments}

%
\appendix

\section{Sign conventions} \label{sec:conventions}

This appendix summarizes our sign conventions. The reader should be careful to sign conventions since various conventions are found in the literature.

Let us consider the response to the conserved current $J^\mu$ under the perturbation of the external source $\calA_\mu$. For simplicity, we assume below that {\it the current $J^\mu$ does not contain the source  $\calA_\mu$.} Suppose that the system is in thermal equilibrium with temperature $1/\beta$ under a static homogeneous external source $\bmcalA_\mu$. Add a perturbation  $\bmcalA_\mu \rightarrow \bmcalA_\mu + \delta \calA_\mu(t, \bmx)$ with the perturbed action
\begin{align}
  \delta S_{\tint}
  := - \int^\beta_0 d\tau \int d\bmx~
  J^\mu(\tau, \bmx)\, \delta \calA_\mu(\tau, \bmx)~.
\label{eq:delta_S}
\end{align}
To linear order in $\delta\calA_\mu$, the deviation of the current expectation value is given by
\begin{align}
   \delta \Exp{J^\mu(\tau, \bmx)}
  &= - \int^\beta_{0} d\tau' \int^\infty_{-\infty} d\bmx'~
  G^{\mu\nu}(\tau - \tau', \bmx - \bmx'\,)\,
\nonumber \\
  &\hspace{1.0truecm}
  \times \delta \calA^\nu(\tau', \bmx'\,)~.
\label{eq:linear_response-app}
\end{align}
Here, $G^{\mu\nu}$ is the Matsubara Green function for the current $J_\mu$:
%
\begin{align}
  & G^{\mu\nu}(\tau, \bmx)
 = - \Exp{ \text{T}_E\, J^\mu(\tau, \bmx)\,
    J^\nu(0, \bm{0}) }_c~,
\label{eq:convention_Matsubara} 
\end{align}
%
where the subscript ``$c$" denotes the connected Green function.
The Fourier transformation of Eq.~(\ref{eq:linear_response-app}) gives
\begin{align}
  & \delta \Exp{\tilJ^\mu(\omega_n, \bmk)}
  = - \tilG^{\mu\nu}(\omega_n, \bmk)\,
  \delta \tcA_\nu(\omega_n, \bmk)~,
\label{eq:FT-linear_response-app}
\end{align}
where $\omega_n = 2 \pi n/\beta$. 

We define the response function $\tilK^{\mu\nu}$ such that it has the same sign as the Green function:
\begin{align}
  & \delta \Exp{\tilJ^\mu(\omega_n, \bmk)}
  = - \tilK^{\mu\nu}(\omega_n, \bmk)\,
  \delta \tcA_\nu(\omega_n, \bmk)~,
\label{eq:convention_response}
\end{align}
%
As is clear from Eqs.~(\ref{eq:FT-linear_response-app}) and (\ref{eq:convention_response}), the response function is nothing but the Matsubara Green function in this case. Accordingly, most literature uses the words response function and Green function interchangeably, but this is the case when the current $J^\mu$ does not contain the source $\calA_\mu$.  When the current does contain the source, they are not the same as we see in the text.

Our response function $\tilK^{\mu\nu}$ differs from the traditional one by a minus sign. It is natural to define the response function $K'$ as $K':= \delta \langle J \rangle/\delta \calA$. Unfortunately,  the response function defined in this way differs from the Green function by a minus sign. 
Our argument in this paper focuses on the sign of these functions, so the traditional convention is somewhat confusing. 

The Matsubara Green function is related to the retarded Green function $G_R$ by analytic continuation:
\be
\tilG^{\mu\nu}_R(\omega) = \left. \tilG^{\mu\nu}(i\omega_n) \right\vert_{i\omega_n=\omega+i0^+}.
\label{eq:convention_analytic}
\ee
The sign convention of the Matsubara Green function (\ref{eq:convention_Matsubara}) is often found in the statistical physics literature whereas the opposite convention is often found in the field theory literature. As a result, in the field theory literature, one needs an extra minus sign in the analytic continuation (\ref{eq:convention_analytic}) as well. We choose the sign convention so that the Matsubara Green function and the retarded Green function have the same sign.

\section{The derivation of the holographic London equation} \label{sec:holo^London_eq}

\subsection{Preliminaries} \label{sec:preliminaries}

For simplicity, we take the ``probe limit" $e\rightarrow\infty$ while keeping $e A_M$ and $e \Psi$ fixed. In this limit, gravity and Maxwell-scalar systems are decoupled, and the problem reduces to the Maxwell-scalar system in the background SAdS$_{p+2}$ spacetime:
\begin{subequations}
\label{eq:metric_p+2}
\begin{align}
  & ds_{p+2}^2
  = \frac{l^2\, \zeta^2}{u^2}\, \left( - f(u)\, dt^2
  + d\bmx^2_p + \frac{du^2}{\zeta^2\, f(u)} \right)~,
\label{eq:line_element} \\
  & f(u) := 1 - u^{p+1}~.
\label{eq:def-f}
\end{align}
\end{subequations}
Here, $\zeta$ is related to the Hawking temperature $T$ as
$4 \pi\, T = (p + 1)\, \zeta$.

From the matter action
\begin{align}
  & S
  = - \int d^{p+2}x~\sqrt{- g}\, \left( \frac{F^{MN}\, F_{MN}}{4}
  + \left\vert\, D \Psi\, \right\vert^2 + V \right)~,
\label{eq:action}
\end{align}
the equations of motion are given by
\begin{subequations}
\label{eq:bg-EOM}
\begin{align}
   0
  &= D^M D_M\, \Psi - V'(|\, \Psi\, |^2)\, \Psi
  ~,
\label{eq:def-scalar_eq} \\
    0
   &= \nabla_N\, F^{MN} - j^M~,
\label{eq:def-EM_eq}
\end{align}
\end{subequations}
where
\begin{align}
   j^M 
  &:= \frac{\delta S_\Psi}{\delta A_M}
  = -i\, e\, g^{MN}\,
    \Big[~\Psi^\dagger\, \big( D_N \Psi \big)
    - \big( D_N \Psi \big)^\dagger \Psi~\Big]~.
\label{eq:def-j}
\end{align}

We focus on an electric solution
$A_M = \bmA_t(u)\, (dt)_M$, 
$\Psi = \bmPsi(u)$
on the SAdS$_{p+2}$ spacetime.
The equations of motion reduce to
\begin{subequations}
\label{eq:bg_eq}
\begin{align}
   0
  &= \left( \odiff{}{u} \frac{f}{u^p} \odiff{}{u}
  - \frac{l^2 m^2}{u^{p+2}}
  + \frac{e^2\, \bmA_t^2}{\zeta^2 u^p f} \right) \bmPsi~,
\label{eq:bg-scalar_eq-I} \\
   0
  &= \left( u^{p-2}
    \odiff{}{u} \frac{1}{ u^{p-2} } \odiff{}{u}
  - \frac{2 l^2\, e^2\, \vert \bmPsi \vert^2}{u^2 f} \right) \bmA_t~,
\label{eq:bg-EM_eq-t-I} \\
   0
  &=\odiff{\bmPsi^\dagger}{u} \bmPsi
  - \bmPsi^\dagger \odiff{\bmPsi}{u}~.
\label{eq:bg-EM_eq-u-I}
\end{align}
\end{subequations}
Equation~(\ref{eq:bg-EM_eq-u-I}) implies that
the phase of $\bmPsi$ must be constant so that
one can set $\bmPsi$ to be real without loss of generality.

The solution to Eqs.~(\ref{eq:bg_eq}) is obtained 
by imposing (i) the regularity condition at the horizon
and (ii) the asymptotically AdS condition.
Condition~(i) is given by
\begin{align}
  & \bmA_t(u=1) = 0~,
& & \bmPsi(u=1) = \text{const.}
\label{eq:bg-bc_horizon}
\end{align}

Condition~(ii) depends on the scalar field mass and thermodynamic ensemble. The asymptotic behavior of matter fields is given by
\begin{subequations}
\label{eq:bg-asymptotic_behavior}
\begin{align}
  & \bmA_t(u)
  \sim \mu - \frac{l\, \rho}{(p - 1) (l\, \zeta)^{p-1}}\, u^{p-1}~,
\label{eq:bg-asymptotic_behavior-I} \\
  & \bmPsi(u)
  \sim \psi^{(-)} u^{\Delta_-} + \psi^{(+)} u^{\Delta_+}~,
\label{eq:bg-asymptotic_behavior-II} \\
  & \Delta_\pm := \frac{p + 1}{2} \pm
  \sqrt{ \left( \frac{p + 1}{2} \right)^2 + l^2 m^2 }~.
\label{eq:def-Delta}
\end{align}
\end{subequations}
Here, $\rho$ is the charge density in the boundary theory, and $\mu$ is the chemical potential associated with the charge density. 
The asymptotic boundary condition for the gauge field depends on which thermodynamic ensemble one is interested in ({\it e.g.}, fixed $\mu$).

For the scalar field, the asymptotic boundary condition depends on its mass:
\begin{subequations}
\label{eq:bg-bc_boundary}
\begin{align}
  & \text{(iia)} \hspace{0.3truecm}
    \psi^{(-)} = 0 \hspace{1.0truecm}
  \text{~~for~~} l^2 m^2 > 1 + l^2 m_{\text{BF}}^2~,
\label{eq:bg-bc_boundary-I} 
\\
  & \text{(iib)} \hspace{0.2truecm}
    \psi^{(-)} = 0 \text{~~or~~} \psi^{(+)} = 0
\nonumber \\
  &\hspace{1.5truecm}
  \text{~~for~~}
  1 + l^2 m^2_{\text{BF}} > l^2 m^2 > l^2 m^2_{\text{BF}}~,
\label{eq:bg-bc_boundary-II}
\end{align}
\end{subequations}
%
%
where $l^2 m^2_{\text{BF}} := - (p+1)^2/4$.
For the case~(\ref{eq:bg-bc_boundary-I}), a normalizable solution must behave as
$\bmPsi \sim u^{\Delta_+}$. 
Thus, $\psi^{(+)}$ represents (the expectation value of) a boundary theory operator $\calO$, and $\psi^{(-)}$ represents the source of the operator. The boundary condition (\ref{eq:bg-bc_boundary-I}) imposes that there is no external source to $\calO$, so the nonvanishing $\psi^{(-)}$ implies a spontaneous condensation of the dual operator. 

For the case~(\ref{eq:bg-bc_boundary-II}), both solutions  $\bmPsi \sim u^{\Delta_-}$ and $\bmPsi \sim u^{\Delta_+}$ are normalizable. Thus, we have two choices for the boundary theory, {\it i.e.}, 
(1)~$\psi^{(+)} \leftrightarrow \text{operator}$ and
$\psi^{(-)} \leftrightarrow \text{source}$, or
(2)~$\psi^{(-)} \leftrightarrow \text{operator}$ and
$\psi^{(+)} \leftrightarrow \text{source}$ \cite{Klebanov:1999tb}.

\subsection{Derivation} 


To derive the holographic London equation, note the following points: 
\begin{itemize}

\item Because the London equation is a static response, one can use the standard Euclidean prescription of the GKP-Witten relation, not the Lorentzian prescription in Ref.~\cite{Son:2002sd}.

\item The London equation describes the property of the transverse part of a static response function (\ref{eq:guess-holographic_London_eq}). Thus, it is enough to consider a static vector perturbation of the Maxwell field. 
The Maxwell perturbations are decomposed into the vector and scalar perturbations. In the frame $k^i=(0,\cdots, 0,k)$, components $A_\alpha$, $\alpha = x_1, \cdots, x_{p-1}$ are the vector perturbations. We consider the perturbations which take the form
\be
\delta A_\alpha = A_\alpha = \tilA_{\alpha, k}(u)\, e^{i k x_{p}}~.
\ee
In the SAdS background, the vector perturbations decouple from the scalar field perturbation $\delta\Psi$ even in the superconducting phase, so one can set $\delta \Psi = 0$.
Also, it is enough to consider the long-wavelength limit for the London equation.
\end{itemize}

From Eq.~(\ref{eq:def-EM_eq}), the ``static" perturbation is given by
\begin{align*}
  & 0
  = \left( u^{p-2}\, \partial_u\, \frac{f}{u^{p-2}}\, \partial_u
  - \frac{k^2}{\zeta^2} - 2\, l^2 e^2\, \frac{|\, \bmPsi\, |^2}{u^2}
  \right) \tilA_{\alpha, k}~,
\end{align*}
where we use the Euclidean version of the metric (\ref{eq:metric_p+2}).
We impose the following boundary conditions: the regularity of 
$\tilA_{\alpha, k}(u \to 1)$ at the horizon, and $\tilA_{\alpha, k}(u \to 0) = \tcA_\alpha(k)$ at the boundary.

Since it is enough to obtain the long-wavelength limit $ k \to 0$, we solve
\begin{subequations}
\label{eq:EOM_system-A_i-probe}
\begin{align}
  & u^{p-2}\, \pdiff{}{u}\, \frac{f}{u^{p-2}}\, \pdiff{}{u}\, \tilA_\alpha
  = 2\, l^2 e^2\, \frac{|\, \bmPsi\, |^2}{u^2}\, \tilA_{\alpha}~,
\label{eq:EOM-A_i-probe} \\
  & \tilA_{\alpha}(u \to 1) = \text{regular}~,
\label{eq:bc_horizon-A_i-probe} \\
  & \tilA_{\alpha}(u \to 0) = \tcA_\alpha(k = 0)
  =: \tcA_\alpha~,
\label{eq:bc_boundary-A_i-probe}
\end{align}
\end{subequations}
where $\tilA_\alpha(u) := \tilA_{\alpha, k=0}(u)$.

Let us consider the Green function $g(u, u'\,)$ which is defined by
\begin{subequations}
\label{eq:def-A_i-Green_func}
\begin{align}
  & \pdiff{}{u}\, \frac{f(u)}{u^{p-2}}\, \pdiff{}{u}\, g(u, u'\,)
  = \delta( u - u'\,)~,
\label{eq:EOM-A_i-Green_func} \\
  & g(u = 0, u'\,) = g(u, u' = 0) = 0~,
\label{eq:bc-A_i-Green_func-bdy} \\
  & g(u = 1, u'\,)\, ,~g(u, u' = 1)~~\text{are regular.}~
\label{eq:bc-A_i-Green_func-horizon}
\end{align}
\end{subequations}
Then, Eq.~(\ref{eq:EOM_system-A_i-probe}) is formally solved as
\begin{subequations}
\label{eq:sol-A_i}
\begin{align}
   \tilA_\alpha(u)
  &= \tcA_\alpha
  + \int^1_0 du'~g(u, u'\,)\, \frac{2\, l^2 e^2}{u'^p}\,
  | \bmPsi(u'\,)\, |^2\, \tilA_\alpha(u'\,)
\nonumber \\
  &= \calF(u)\, \tcA_\alpha
  + O\left( | \bmPsi\, |^4 \right)~,
\label{eq:sol-A_i-probe} \\
   \calF(u)
  &:= 1 + \int^1_0 du'~g(u, u'\,)\, \frac{2\, l^2 e^2}{u'^p}\,
  | \bmPsi(u'\,)\, |^2~.
\label{eq:def-calF}
\end{align}
\end{subequations}

From Eq.~(\ref{eq:action}), one obtains the (Euclidean) on-shell action for the perturbation:
\begin{align*}
   S^{V}_{\text{os}}
  &= - \left. \frac{\zeta (l \zeta)^{p-2}}{2 (2 \pi)^{p+1}}\,
  \delta^{\alpha\beta}\, \tilA_\alpha\,
  \frac{f(u)}{u^{p-2}}\, \partial_u \tilA_\beta\, \right\vert_{u=0}
  + O( k^2 )~.
\end{align*}
Using Eq.~(\ref{eq:def-R_current-response_func}), the static response function $\tilK^{\alpha\beta}(k)$ of the current is given by
\bea
   \lim_{k \to 0} \tilK^{\alpha\beta}(k)
  &=& \left. - \zeta (l \zeta)^{p-2}\, \delta^{\alpha\beta}\,
  \frac{f(u)}{u^{p-2}}\, \partial_u \calF(u)\, \right\vert_{u=0}
\nonumber \\
  &&+ O\left( | \bmPsi\, |^4 \right)~
\label{eq:R_current-response_func-pre}
\eea
in the long-wavelength limit. 

Now, from Eqs.~(\ref{eq:EOM-A_i-Green_func}) and
(\ref{eq:bc-A_i-Green_func-horizon}),
\begin{align}
  & \frac{f(u)}{u^{p-2}}\, \pdiff{}{u}\, g(u, u'\,)
  = - \theta( u' - u )~,
\label{eq:A_i-Green_func-prime}
\end{align}
so
\begin{align*}
   \frac{f}{u^{p-2}}\, \partial_u \calF
  &= - 2\, l^2 e^2\,
  \int^1_u du'~\frac{| \bmPsi(u'\,)\, |^2}{u'^p}~.
\end{align*}
Using the above equation, one obtains the final expression of the static response function for the current in the long-wavelength limit:
\bea
\lim_{k \to 0} \tilK^{\alpha\beta}(k)
  &=& 2\, (l^2 e^2)\, \zeta\, (l \zeta)^{p-2}\, \delta^{\alpha\beta}\,
  \int^1_0 du~\frac{| \bmPsi\, |^2}{u^p}
\nonumber \\
  &&+ O\left( | \bmPsi\, |^4 \right)~.
\label{eq:R_current-response_func}
\eea
The electric background we consider causes the second-order phase transition, and 
$|\bmPsi |$ is small enough near the critical point, so Eq.~(\ref{eq:R_current-response_func}) becomes a good approximation near the critical point. 

Equation~(\ref{eq:R_current-response_func}) corresponds to the transverse part of the response function. 
Since $| \bmPsi | \propto \Exp{\calO}$, 
\begin{align}
   \lim_{\bmk \to 0} \tilK_{T}(\bmk)
  &\sim (\text{positive const.}) \times | \Exp{\calO} |^2~.
\label{eq:R_current-response_func-near_Tc}
\end{align}
Therefore, Eq.~(\ref{eq:guess-holographic_London_eq}) holds with the natural correspondence $n_s \sim | \Exp{\calO} |^2$.

The above result (\ref{eq:R_current-response_func-near_Tc}) holds irrespective of the spatial dimensionality $p$, scalar field mass $m$, and its boundary condition as long as the transition from $\bmPsi = 0$ to $\bmPsi \ne 0$ is second-order.

\section{Holographic renormalization} \label{sec:counter_term}

When one evaluates the on-shell action, the counterterm action is often necessary. But there is no divergence for the $U(1)$ current, so we ignore the counterterms in \appen{holo^London_eq}. In this Appendix, we examine if the counterterms do not induce finite renormalizations.

The counterterm action has not been derived for the Einstein-Maxwell-complex scalar system.%
\footnote{
If one evaluates the counterterm action using the Hamilton-Jacobi method, one naively gets no solution.
}
We evaluate the counterterm action from the Einstein-Maxwell-real scalar system.

The Euclidean action of the Einstein-Maxwell-real scalar system is given by
\begin{align}
  & S
  = \int_\calM d^{p+2}x~\sqrt{g}~\bigg( - R
  + 2 \Lambda + \frac{ F^{MN} F_{MN} }{4}
\nonumber \\
  &\hspace{3.0truecm}
  + \frac{( \nabla \Phi )^2}{2}
  + \frac{m^2}{2}\, \Phi^2 + \cdots \bigg)~,
\label{eq:E-M-real_scalar-L}
\end{align}
The counterterm action is given by \cite{Batrachenko:2004fd}%
\footnote{In addition to the counterterm action, one also needs the Gibbons-Hawking term, but it does not contribute to the Maxwell field we are interested in.
}
\begin{subequations}
\label{eq:E-M-real_scalar-S_ct}
\begin{align}
  & S_\tct
  = \int_{\partial \calM} d^{p+1}x~\sqrt{\gamma}
\nonumber \\
  &\times \bigg[~\frac{2 p}{l} + C(\Phi)\, \calR
  + \frac{M(\Phi)}{4}\, \gamma^{\mu\rho} \gamma^{\nu\sigma}\,
    F_{\mu\nu} F_{\rho\sigma}
\nonumber \\
  &+ \frac{\calM(\Phi)}{2}\, \, \gamma^{\mu\nu}
    (\partial_\mu \Phi) (\partial_\nu \Phi)
  + \frac{\Delta_-}{2 l}\, \Phi^2
  + \cdots \bigg]~,
\label{eq:E-M-real_scalar-L_ct} \\
  & C(\Phi)
  = \frac{l}{p - 1}
  - \frac{l}{4 p}\, \frac{\Delta_-}{\Delta_+ - \Delta_- - 2}\, \Phi^2
  + \cdots~,
\label{eq:real_scalar-coeff_C-L_ct} \\
  & M(\Phi)
  = - \frac{l}{p - 3} + \cdots~,
\label{eq:real_scalar-coeff_M-L_ct} \\
  & \calM(\Phi)
  = - \frac{l}{\Delta_+ - \Delta_- - 2} + \cdots~.
\label{eq:real_scalar-coeff_calM-L_ct}
\end{align}
\end{subequations}
Here, $\gamma_{\mu\nu}$ and $\calR$ are the induced metric and its Ricci scalar on the AdS boundary $\partial \calM$, respectively. The coefficients $\Delta_\pm$ are determined from the scalar mass $m$ in Eq.~(\ref{eq:def-Delta}). The counterterm action (\ref{eq:E-M-real_scalar-S_ct}) is valid for $p \le 3$ and $m_\tBF^2 < m^2 < (3/4) m_\tBF^2$.

The counterterm action for the Einstein-Maxwell-complex scalar system can be obtained by (1) replacing the real scalar $\Phi$ by the complex scalar $\sqrt{2}\, \Psi$, and by (2) replacing the derivative by the gauge covariant derivative 
$\partial_M \to D_M = \partial_M - i e\, A_M$:
\begin{subequations}
\label{eq:E-M-scalar-S_ct}
\begin{align}
  & S_\tct
  = \int_{\partial \calM} d^{p+1}x~\sqrt{\gamma}
\nonumber \\
  &\times \bigg[~\frac{2 p}{l} + C(\Psi)\, \calR
  + \frac{M(\Psi)}{4}\, \gamma^{\mu\rho} \gamma^{\nu\sigma}\,
    F_{\mu\nu} F_{\rho\sigma}
\nonumber \\
  &+ \calM(\Psi)\, \gamma^{\mu\nu} ( D_\mu \Psi )^\dagger\,
    ( D_\nu \Psi )
  + \frac{\Delta_-}{l}\, |\Psi |^2
  + \cdots \bigg]~,
\label{eq:E-M-scalar-L_ct} \\
  & C(\Psi)
  = \frac{l}{p - 1}
  - \frac{l}{2 p}\, \frac{\Delta_-}{\Delta_+ - \Delta_- - 2}\,
    | \Psi |^2 + \cdots~,
\label{eq:coeff_C-L_ct} \\
  & M(\Psi)
  = - \frac{l}{p - 3} + \cdots~,
\label{eq:coeff_M-L_ct} \\
  & \calM(\Psi)
  = - \frac{l}{\Delta_+ - \Delta_- - 2} + \cdots~.
\label{eq:coeff_calM-L_ct}
\end{align}
\end{subequations}
The static response function for the $U(1)$ current may have contributions from the counterterm action (\ref{eq:E-M-scalar-L_ct}).

The contributions could come from $F_{\mu\nu}$ and
$(\partial_\mu - i e\, A_\mu) \Psi$.
The former does not contribute in the static homogeneous limit. Let us examine the latter for the vector perturbation. When $\psi^{(+)}$ has a spontaneous condensation ($\bmPsi \sim \psi^{(+)}\, u^{\Delta_+}$),
%
%
\bea
   S_\tct
  &=& - (l\, \zeta)^{p-1}\, e^2 \delta^{\alpha\beta} \calA_\alpha \calA_\beta
  \calM\big( \psi^{(+)} \big)\,
  \big\vert \psi^{(+)}  \big\vert^2\,
\nonumber \\
  & &\times u^{\Delta_+ - \Delta_-}\, \Big\vert_{u=0}
  = 0~,
\label{eq:E-M-scalar-L_ct-for_vector}
\eea
so the counterterm action again makes no contribution to the response function. 

Therefore, the static response function of the $U(1)$ current has no contribution from the counterterm action (\ref{eq:E-M-scalar-S_ct}), and Eq.~(\ref{eq:def-R_current-response_func}) remains valid.

\section{The positivity of spectral function}
\label{sec:negativity}

First, consider the Wightman function:
\begin{align}
   S^{\mu\nu}(x)
  &:= \Exp{ J^\mu(x)\, J^\nu(0) }~,
\label{eq:def-calS_munu}
\end{align}
where $x = (t, \bmx)$. From the Hermiticity of $J^\mu$ and translational invariance, 
\begin{subequations}
\begin{align}
  \big[\, S^{\mu\nu}(x)\, \big]^* &= S^{\nu\mu}(- x)~,
\label{eq:prop-calS_munu} \\
  \Leftrightarrow \hspace{0.2truecm}
  \big[\, \tilS^{\mu\nu}(k)\, \big]^* &= \tilS^{\nu\mu}(k)~,
\label{eq:prop-tcS_munu}
\end{align}
\end{subequations}
where $k = (\omega, \bmk)$. 

Using the translation operator $P^\mu$ ($P \cdot x := P_\mu x^\mu$), one obtains
\begin{align}
  & \tilS^{\mu\nu}(k)
  = \int^\infty_{-\infty} d^{p+1}x~e^{- i k \cdot x}
  \Exp{ e^{- i P \cdot x}\, J^\mu(0)~
        e^{i P \cdot x}\, J^\nu(0) }
\nonumber \\
  &= \int^\infty_{-\infty} d^{p+1}x~\frac{e^{- i k \cdot x}}{Z(\beta)}~
  \Tr\left[\, e^{-\beta P^0}
    e^{- i P \cdot x} J^\mu(0)\, e^{i P \cdot x} J^\nu(0)
     \right]
\nonumber \\
  &= \frac{(2 \pi)^{p+1}}{Z(\beta)}
  \sum_{n, n'} e^{- \beta q_n^0}\, \delta( q_{n'} - q_n - k )
\nonumber \\
  &\hspace*{1.5truecm}
  \times \bra{n} J^\mu(0) \ket{n'} \bra{n'} J^\nu(0) \ket{n}
\nonumber \\
  &= e^{\beta k^0}\, \tilS^{\nu\mu}(-k)~,
\label{eq:tcS_munu}
\end{align}
from which one can derive the fluctuation-dissipation theorem. 
Here, $Z(\beta) := \Tr( e^{- \beta P^0} )$, and 
$\ket{n}$ forms a complete eigensystem of $P^\mu$:
\begin{align}
  & 1 = \sum_n \ket{n}\bra{n}~,
& & P^\mu \ket{n} = q^\mu_n \ket{n}~.
\label{eq:eigen_system-P}
\end{align}

In terms of the Wightman function $S^{\mu\nu}$, the spectral function is written as 
\begin{subequations}
\label{eq:Delta-calS}
\begin{align}
   \rho^{\mu\nu}(x)
  &= S^{\mu\nu}(x) - S^{\nu\mu}(-x)~,
\label{eq:Delta_munu-calS_munu} \\
  \Leftrightarrow \hspace{0.2truecm}
   \tilrho^{\mu\nu}(k)
  &= \tilS^{\mu\nu}(k) - \tilS^{\nu\mu}(-k)~.
\label{eq:tilDelta_munu-tcS_munu}
\end{align}
\end{subequations}
%
From \eq{prop-tcS_munu}, 
\begin{align}
  & \big[\, \tilrho^{\mu\nu}(k)\, \big]^*
  = \tilrho^{\nu\mu}(k)~.
\label{eq:prop-tilDelta-app}
\end{align}
In particular, $\tilrho^{\mu\mu}(k)$ is real. 

Also, \eq{tcS_munu} leads to
\begin{align}
  & \tilrho^{\mu\nu}(\omega, \bmk)
  = ( 1 - e^{- \beta \omega} )\, \tilS^{\mu\nu}(\omega, \bmk)~.
\label{eq:tilDelta_munu-tcS_munu-II}
\end{align}
For diagonal components,
\begin{align*}
  & \tilS^{\mu\mu}(k)
  = \frac{(2 \pi)^{p+1}}{Z(\beta)}
  \sum_{n, n'} e^{- \beta q_n^0}\, \delta( q_{n'} - q_n - k )
\nonumber \\
  &\hspace*{1.5truecm}
  \times \left\vert\, \bra{n} J^\mu(0) \ket{n'}\, \right\vert^2
  > 0~,
\end{align*}
so one obtains
\begin{align}
  & \frac{\tilrho^{\mu\mu}(\omega, \bmk)}{\omega}
  = \frac{\sinh(\beta \omega/2)}{\omega}\,
  e^{- \beta \omega/2}\, \tilS^{\mu\mu}(\omega, \bmk)
  > 0~.
\label{eq:negativity-app}
\end{align}
\vspace{0.2truecm}


\end{document}